\documentclass[reprint,groupedaddress,amsmath,amssymb,aps,showkeys]{revtex4-1}

\usepackage{graphicx}
\usepackage{dcolumn}
\usepackage{bm}
\usepackage{hyperref}
\usepackage{physics}
\usepackage{mathtools}
\usepackage{subcaption}

\allowdisplaybreaks

\def\p{\partial}

\usepackage{graphicx,amssymb,amsmath,amsthm,amsfonts}
\usepackage[usenames]{color}
\usepackage{xcolor}
\usepackage{mathtools}

\begin{document}

\title{Normal modes of Proca fields in AdS
  spacetime}

\author{Tiago V. Fernandes}
\affiliation{Centro de Astrof\'{\i}sica e Gravita\c c\~ao  - CENTRA,
Departamento de F\'{\i}sica, Instituto Superior T\'ecnico - IST,
Universidade de Lisboa - UL,
Avenida Rovisco Pais 1, 1049-001 Lisboa, Portugal}
\author{David Hilditch}
\affiliation{Centro de Astrof\'{\i}sica e Gravita\c c\~ao  - CENTRA,
Departamento de F\'{\i}sica, Instituto Superior T\'ecnico - IST,
Universidade de Lisboa - UL,
Avenida Rovisco Pais 1, 1049-001 Lisboa, Portugal}
\author{Jos\'{e} P. S. Lemos}
\affiliation{Centro de Astrof\'{\i}sica e Gravita\c c\~ao  - CENTRA,
Departamento de F\'{\i}sica, Instituto Superior T\'ecnico - IST,
Universidade de Lisboa - UL,
Avenida Rovisco Pais 1, 1049-001 Lisboa, Portugal}
\author{V\'itor Cardoso}
\affiliation{Centro de Astrof\'{\i}sica e Gravita\c c\~ao  - CENTRA,
Departamento de F\'{\i}sica, Instituto Superior T\'ecnico - IST,
Universidade de Lisboa - UL,
Avenida Rovisco Pais 1, 1049-001 Lisboa, Portugal}


\begin{abstract}
A normal  mode analysis for Proca fields in the  anti-de Sitter
(AdS) spacetime is given. It is found that the equations for the Proca
field can be decoupled analytically. This is performed by changing the
basis of the vector spherical harmonics (VSH) decomposition.  The
normal modes and the normal mode frequencies of the Proca equation in
the AdS spacetime are then analytically determined.
It is also shown that the
Maxwell field can be recovered by taking the massless limit of the
Proca field with care so that the nonphysical gauge modes are
eliminated.
\end{abstract}


\maketitle

\section{Introduction}\label{sec:Intro}

The anti-de Sitter (AdS)
spacetime~\cite{Penrose:1967,hawking_ellis_1973} is a vacuum
solution to the Einstein equations with negative cosmological
constant, and has therefore negative scalar curvature.  There has been
substantial interest in studying AdS spacetimes for several
reasons. One is that their asymptotic behavior is well controlled and
permits further analytical treatment of the spacetime
\cite{isham1978}.  Another is due
to the bridge between asymptotically AdS spacetimes and conformal
field theories that naturally arise in string theory, a link known as
the AdS/CFT conjecture~\cite{Maldacena:1997}.  In addition, since the
AdS spacetime is a confined spacetime, it has been used to emulate the
dynamics of fields in a cavity.  In this connection, a gravitational
instability of AdS spacetime has been found in~\cite{Bizon:2011gg}
where, by considering the numerical evolution of the
Einstein--Klein-Gordon equations, it was shown that nonlinear effects
always become important at late times and lead to the formation of
small black holes from a large class of arbitrary small initial
amplitudes of the scalar field. Further work supports this claim by
considering a complex scalar field in
AdS~\cite{Buchel:2012}. Therefore, the AdS spacetime is unstable to
certain types of initial data, although not against all classes of
initial data, see, e.g.,~\cite{masachsway}.

An important aspect of many spacetimes is the behavior of their
quasinormal
and normal modes, as these modes describe how the spacetimes
respond under linear perturbations.
In asymptotically AdS spacetimes,
for which the oscillations are quasinormal,
these modes
play a role in the linear stability of the spacetimes
themselves and have also proven of interest in
the AdS/CFT conjecture.
In the  AdS
spacetime, one finds that 
when it is linearly perturbed with
reflective boundary conditions, also called box
boundary conditions,
it oscillates normally, rather than quasinormally,
as there are no losses of the field to infinity.
For the  AdS spacetime, the normal modes of
scalar~\cite{Burgess:1984ti}, electromagnetic and gravitational
perturbations \cite{Cardoso:2003cj}, and massive Proca
perturbations~\cite{Konoplya:2006,ovsiyuk2011spherical} have been
studied.

In this work, we obtain analytically the normal modes and the normal
mode frequencies of the Proca field in the  AdS spacetime under
reflective boundary conditions.  To do so we perform, in a new way, a
change of the basis of the usual vector spherical harmonics (VSH)
decomposition that not only separates the Proca equations but also
decouples the relevant modes, thus remarkably simplifying the
calculations.  The Maxwell field normal modes and normal mode
frequencies can be recovered distinctly by taking with care the
massless limit of the Proca field.  We use some methods and results
taken from~\cite{Ruffini:1973,Rosa:2012,tfprocasch,Sakurai:2017,
Abramowitz:1964}.

The paper is organized as follows. In
section~\ref{sec:EinsteinProcafieldequations}, we introduce the
classical treatment of minimally coupled Proca field to the curvature
and display the AdS line element
for the fixed background. In section~\ref{sec:ProcafieldADS},
the separation and decoupling of the Proca field in AdS spacetime
through the VSH method is studied. In section~\ref{sec:normalmodes},
imposing reflective boundary conditions for the Proca field, 
we calculate analytically the Proca normal modes and normal mode
frequencies, and find also the Maxwell modes by taking the zero mass
limit. In section~\ref{sec:Conc}, we conclude. In
the Appendices \ref{notation} and \ref{derivation}
we give formulas and perform
some calculations that are useful for the main text.
We use geometric units
throughout the paper, so that~$G = c = 1$.  

\section{Field equations for the Einstein-Proca system with a negative
  cosmological constant: Proca fields in AdS spacetime }
\label{sec:EinsteinProcafieldequations}

\subsection{Proca field equations}

The Einstein-Proca system is described by the action $ S = \int
\dd[4]x \sqrt{-g} \left(L_{\rm EH} - L_{\rm P}\right) $, where $g$ is
the determinant of the metric $g_{ab}$, $ L_{\rm EH}=\frac{R -
  2\Lambda}{16\pi} $, is the Einstein-Hilbert Lagrangian density,
  with
$R_{ab}$ and $R = R_{ab}g^{ab}$ being the Ricci tensor and the Ricci
scalar, respectively, which
in turn are related to the metric $g_{ab}$ and its
first and second spacetime derivatives, $\Lambda$
is the cosmological
constant, $ L_{\rm P} = \frac{F_{ab}F^{ab}}{4} + \frac{\mu^2}{2} A_a
A^a $ is the Proca Lagrangian density,
with $F_{ab} = \nabla_{a} A_{b} -
\nabla_{b} A_{a}$ being the Proca field strength, $\nabla_{a}$
denoting
covariant derivative, $A_a$
being the Proca vector potential, and $\mu$ being 
the mass of the Proca field. The Euler-Lagrange equations for the
action above yield the Einstein equation $ G_{ab} + \Lambda g_{ab} =
8\pi T_{ab} $ for the metric
field, where $T_{ab}$ is the Proca energy-momentum tensor
given by $ T_{ab} = F_{ae}F_{bf}g^{ef} + \mu^2 A_a A_b - g_{ab}L_P $,
and the Proca equation $\nabla_b F^{ab} + \mu^2 A^a = 0$ for the
vector field $A_a$. There are internal equations for $A_a$, namely,
$\nabla_{[a}F_{bc]} = 0$. Moreover,
as a direct
consequence of the action formalism, the metric field and the
Proca field obey their
own Bianchi identities, namely,
$\nabla_a (G^{ab} + \Lambda g^{ab})=0$, which implies 
$\nabla_a T^{ab} = 0$, and
$\nabla_aA^a = 0$, respectively.
We are interested in
small, first order,
perturbations of a Proca vector field $A_a$
on a fixed
spacetime background with negative cosmological constant
$\Lambda$, i.e., $\Lambda<0$.

In zeroth
order in $A_a$, i.e., for zero  Proca vector field,
the Einstein equation reduces to
$G_{ab} - \frac{3}{l^2}g_{ab}=0$, 
where $l= \sqrt{-\frac{3}{\Lambda}}$ is
the spacetime
characteristic length
associated to the negative
cosmological constant $\Lambda$.
This Einstein equation 
gives
as a solution 
the  AdS spacetime,
which is here considered as
the fixed background metric.
The AdS spacetime
is  spherically
symmetric, has  no horizons,
and its line element in
spherical 
coordinates $(t,r,\theta,\phi)$
is
given by
\begin{align}
  ds^2 =& - f(r) dt^2 + \dfrac{dr^2}{f(r)}
  + r^2 (d\theta^2 + \sin^2\theta d\phi^2)\,,
\nonumber\\
&f(r) = 1 +\frac{r^2}{l^2}\,.
\label{eq:fSchwADS}
\end{align}

In first order in $A_a$, i.e., for perturbations of the field $A_a$
that generate small curvature perturbations compared with the
curvature of the background, it is possible to perform an analytical
treatment.  In first order in $A_a$, there is a corresponding Proca
energy-momentum tensor $T_{ab}$.  This Proca energy-momentum tensor
$T_{ab}$ is second order in $A_a$, which means that it does not back
react in first order into the vacuum Einstein equation, i.e., $G_{ab}
+ \frac{3}{l^2}g_{ab}=0$, in turn implying that the AdS line element
of Eq.~\eqref{eq:fSchwADS} still holds.
It also means that  
the Proca
$T_{ab}$ 
still
obeys the Bianchi identity
$\nabla_a T^{ab} = 0$,
on
the fixed AdS background metric
solution of Eq.~\eqref{eq:fSchwADS}. 
On the other hand, since
now there is a first order
Proca field $A_a$
permeating the spacetime, the field 
has to obey in first order
the Proca equation $\nabla_b F^{ab} +
\mu^2 A^a = 0$. This equation
can be simplified to
\begin{align}
  g^{cd}\nabla_c\nabla_d A^a
  - \Big(\mu^2 - \frac{3}{l^2}\Big)A^a = 0 \,,
  \label{eq:Procaeqvacuum}
\end{align}
where 
the commutator property
of the covariant derivative has been used,
and where it was imposed 
the Bianchi
identity for the field $A_a$, i.e., 
the
divergence of the vector field
$A^a$ is zero, 
explicitly
\begin{align}
  \nabla_aA^a = 0\,.
  \label{eq:LorenzCondition}
\end{align}
The Bianchi identity is a
consequence of the field equations, and
as such is a condition that the field
$A_a$ necessarily obeys.

In brief, first order perturbations of the Proca field $A_a$ in the
AdS spacetime background of Eq.~\eqref{eq:fSchwADS} are described by
Eq.~\eqref{eq:Procaeqvacuum}, with Eq.~\eqref{eq:LorenzCondition} as
an auxiliary equation.

\subsection{Maxwell field equations
as the zero mass limit of the Proca field}

The equations for the first order perturbations of the Maxwell field
in the AdS spacetime background of Eq.~\eqref{eq:fSchwADS} can be
taken as the zero mass limit,~$\mu=0$, of the Proca field equations.
Thus, the massless case~$\mu=0$ reduces Eq.~\eqref{eq:Procaeqvacuum}
to the Maxwell equation for the Maxwell vector field $A_a$, namely
\begin{align}
  g^{cd}\nabla_c\nabla_d A^a
  + \frac{3}{l^2}\,A^a = 0 \,,
  \label{eq:Maxwelleqvacuum}
\end{align}
and the Bianchi identity, Eq.~\eqref{eq:LorenzCondition}, turns into
the Lorenz gauge condition, i.e., the field $A_a$ has not necessarily
to obey it.  In addition, when this gauge condition is imposed, the
Maxwell field still has a residual gauge freedom and can be
transformed as $A^a \rightarrow A^a + \chi^{,a}$, where~$\chi$ is a
scalar field that obeys the Klein-Gordon equation, while the Maxwell
equation remains unchanged.

\section{Proca field perturbations in  AdS spacetime}
\label{sec:ProcafieldADS}

\subsection{Perturbations in the Proca field
and separation of variables}

The Proca equations~\eqref{eq:Procaeqvacuum} in AdS spacetime may be
separated using the VSH method which has been used in several other
instances~\cite{Ruffini:1973,Rosa:2012,tfprocasch}. The VSH method can
be formulated by the sum of a spin~$s=1$ with the angular
momentum~$\ell$ as is well known from quantum
mechanics~\cite{Sakurai:2017}. And so, the following ansatz can be
assumed for the field $A_a$,
\begin{align}
  A_{a} = \frac{1}{r}\sum_{i=0}^3\sum_{l m} q_{(i)}(t,r)
  Q_{(i)\,a}^{\;\ell m}(r,\theta,\phi)\,,
  \label{eq:SchwParamq}
\end{align}
where the $Q_{(i)\,a}^{\;\ell m}$ are given by
\begin{align}
    Q_{(0)\,a}^{\;\ell m} &= (1,0,0,0) Y^{\ell m}\,,\nonumber\\
    Q_{(1)\,a}^{\;\ell m} &= \dfrac1\ell\left(0,\dfrac\ell{rf},
\p_{\theta},\p_{\phi}\right)Y^{\ell m}\,,\nonumber\\
    Q_{(2)\,a}^{\;\ell m} &=  \dfrac1{\ell+1}\left(0,
\dfrac{\ell+1}{rf},-\p_{\theta},-\p_{\phi}\right)Y^{\ell m}\,,
\nonumber\\
    Q_{(3)\,a}^{\;\ell m} &= \frac{r}{\sqrt{\ell(\ell+1)}}
    \Bigg(0,0,\frac{\p_{\phi}}{\sin\theta},
    -\sin\theta\p_{\theta}\Bigg)Y^{\ell m}\,,
\label{Qs}
\end{align}
$Y^{\ell m}$ are the spherical harmonics, $\ell$ is the
angular momentum
number, $m$ is the azimuthal number, and the~$q_{(i)}(t,r)$ should
have been written as~$q_{(i)}^{\ell m}(t,r)$, $q_{(i)}(t,r)\equiv
q_{(i)}^{\ell m}(t,r)$, but to not overcrowd the notation we dropped
the indices~$\ell$ and~$m$ here. For aspects on the notation see
Appendix~\ref{notation}.

Let us define the radius~$r_*$ by~$\dv{r_*}{r} =
\frac1f$ and the family
of operators~$\hat{\mathcal{D}}_p$ by
\begin{equation}
  \hat{\mathcal{D}}_p = - \p_t^2 + \p_{r_*}^2
  - f\Big[\frac{p(p+1)}{r^2} + \mu^2\Big]\,,
  \label{D}
\end{equation}
for a certain parameter~$p$. Then, by inserting the ansatz given in
Eq.~(\ref{eq:SchwParamq}) into the Proca
equations~\eqref{eq:Procaeqvacuum}, one obtains after rearrangements
the following system of equations for the functions~$q_{(i)}$
\begin{align}
  &\hat{\mathcal{D}}_\ell q_{(0)} +
  (\p_r f)(\dot{q}_{(1)} + \dot{q}_{(2)} - q'_{(0)}) = 0 \,,
  \label{eq:SchwEM1}\\
  &\hat{\mathcal{D}}_{j_1} q_{(1)} = 0
    \,,\label{eq:SchwEMi}
\\
  &\hat{\mathcal{D}}_{j_2} q_{(2)} = 0
    \,,\label{eq:SchwEMi2}
\\
  &\hat{\mathcal{D}}_{j_3} q_{(3)} = 0
    \,,\label{eq:SchwEMi3}
\end{align}
where $j_1 = \ell -1$, $j_2 = \ell + 1$, $j_3 = \ell$,~$\dot{q}_{(i)}
= \frac {\partial{q_{(i)}}} {\partial t}$, and $q'_{(i)} = \frac
{\partial{q_{(i)}}} {\partial r_*}$.  The functions $q_{(0)}$,
$q_{(1)}$ and $q_{(2)}$ describe the electric modes and $q_{(3)}$
describes the magnetic modes.  These definitions come from the fact
that the $Q_{(0)\,a}^{\;\ell m}$, $Q_{(1)\,a}^{\;\ell m}$, and
$Q_{(2)\,a}^{\;\ell m}$, gain a factor of~$(-1)^{\ell}$, and the
$Q_{(3)\,a}^{\;\ell m}$ gain a factor of~$(-1)^{\ell+1}$, under parity
transformation.  It must be also noted that the subscript of the
operator $\hat{\mathcal{D}}_p$ in Eq.~\eqref{D} determines the value
of the term appearing in~$\frac{f}{r^2}$, e.g.,
$\hat{\mathcal{D}}_\ell = - \p_t^2 + \p_{r_*}^2 -
f\Big[\frac{\ell(\ell+1)}{r^2} + \mu^2\Big]$.

The Bianchi identity Eq.~(\ref{eq:LorenzCondition}), $\nabla^a A_a
=0$, with the ansatz given by Eq.~\eqref{eq:SchwParamq}, can be
written as
\begin{equation}
  \frac{1}{r f}\Bigg[ - \dot{q}_{(0)}+    q'_{(1)} + q'_{(2)}  
  + \frac{f}{r} \left(- \ell q_{(1)}+(\ell + 1)q_{(2)} 
   \right) \Bigg] = 0\,,
  \label{eq:LorenzConditionSchwq}
\end{equation}
which was used to find Eqs.~(\ref{eq:SchwEMi})
and~(\ref{eq:SchwEMi2}). Indeed, by placing the ansatz
Eq.~(\ref{eq:SchwParamq}) into Eq.~\eqref{eq:Procaeqvacuum}, one
obtains directly $\hat{\mathcal{D}}_\ell q_{(1)} +
\hat{\mathcal{D}}_\ell q_{(2)} + (\p_r f) (\dot{q}_{(0)} - q'_{(1)} -
q'_{(2)}) + \frac{2 f^2}{r^2}(\ell q_{(1)} - (\ell + 1)q_{(2)}) = 0$
and $(\ell +
1)\hat{\mathcal{D}}_{\ell}q_{(1)}-\ell\hat{\mathcal{D}}_{\ell}q_{(2)}
+ \frac{2f\ell(\ell+1)}{r^2}(q_{(1)} + q_{(2)}) = 0$. Substituting
$\dot{q}_{(0)}$ in the two equations above by
Eq.~(\ref{eq:LorenzConditionSchwq}) and further rearranging them,
Eqs.~(\ref{eq:SchwEMi}) and~(\ref{eq:SchwEMi2}) can be obtained.
Equation~(\ref{eq:SchwEMi3}) is obtained directly from the Proca
equation~\eqref{eq:Procaeqvacuum} with the ansatz
Eq.~(\ref{eq:SchwParamq}).  Thus, the system consisting of second
order partial differential equations given in
Eqs.~\eqref{eq:SchwEM1}-\eqref{eq:SchwEMi3} determine the solution of
the Proca field under our ansatz. We can use the Bianchi identity
equation~\eqref{eq:LorenzConditionSchwq} to help in the determination
of the solution, as it is a first order partial differential equation.

To deal with~$q_{(0)}$ we can split every~$q_{(i)}$, $i=0,1,2,3$, into
a static part~$q_{(i)\text{s}}(r)$ and a dynamic part~$q_{(i)\text{d}}
(r,t)$, i.e. $q_{(i)} = q_{(i)\text{s}}(r) +
q_{(i)\text{d}}(t,r)$. Here, we are interested only on the dynamic
solutions of the system. Thus, we can use the Bianchi identity given
in Eq.~(\ref{eq:LorenzConditionSchwq}) in the form~$
{\dot{q}}_{(0)\text{d}} = q'_{(1)} + q'_{(2)} + \frac{f}{r} \left( -
\ell q_{(1)}+ (\ell + 1) q_{(2)} \right) $, to determine the dynamic
part of $q_{(0)}$. Dealing with with~$q_{(1)}$, $q_{(2)}$, and
$q_{(3)}$ is simpler, since the corresponding equations of the system
are completely decoupled. To each~$q_{(i)}$, $i=1,2,3$, there is
a~$j_i$, the total angular momentum, i.e., the angular momentum plus
the spin of the vector. The~$j_1$ cannot be negative, $j_1 \geq 0$,
and moreover, $j_2 , j_3 \geq 1$. For the monopole case, $\ell = 0$,
the only function that is relevant is~$q_{(2)}$.

\subsection{Perturbations in the Maxwell field
as the zero mass limit of the Proca field}

In the massless $\mu=0$ case, one has that
the Proca field equation  for
$A_a$ given in Eq.~\eqref{eq:Procaeqvacuum}
turns into the
Maxwell field equation for
$A_a$ given in Eq.~\eqref{eq:Maxwelleqvacuum}. 
Moreover, in this case 
Eq.~\eqref{eq:LorenzCondition} turns
from a Bianchi identity equation into
the Lorenz gauge condition.  But even imposing
$\nabla_a A^a=0$ as a gauge
in the
massless case, there is
still a further gauge freedom.  This means
that the vector potential $A^a$
includes in this case degrees of freedom that are nonphysical.  Thus,
it is useful in this situation to work with the Maxwell
tensor~$F_{ab}$ rather than~$A_a$. Then, the vector field will have
two physical degrees of freedom corresponding to the two transversal
polarizations that will be described by
\begin{eqnarray}
    &q_{(12)}(r) &= -f
    \left[\frac{\partial_\theta(\sin(\theta)F_{r\theta})}{\sin(\theta)}
    + \frac{1}{\sin^2(\theta)}\partial_\phi F_{r\phi}\right]\,\,,
    \label{q12Max}\\
    &q_{(3)}(r) &= -\frac{F_{\theta \phi}}{\sin(\theta)Y^{\ell
    m}}\,\,,    
    \label{q3Max}
\end{eqnarray}
where $q_{(12)}(r) = - \ell q'_{(1)} +(\ell + 1)q'_{(2)} -
\frac{\ell(\ell + 1)}{r}(q_{(1)} + q_{(2)})$.  The equation for
$q_{(12)}(r)$ can be found as $\hat{\mathcal{D}}_{\ell}q_{(12)}(r) =
0$ by using Eqs.~\eqref{eq:SchwEMi} and \eqref{eq:SchwEMi2}, thus
$q_{(12)}(r)$ will obey the same equation for $q_{(3)}$ with $\mu=0$,
see Eq.~\eqref{eq:SchwEMi3}.  The equation for $q_{(3)}$ will be
Eq.~\eqref{eq:SchwEMi3} with $\mu=0$.

\section{Normal modes of Proca  fields in
 AdS spacetime\label{sec:normalmodes}}

\subsection{Normal modes of the Proca field}

The normal modes of the Proca field in an AdS spacetime are described
by the solutions of Eqs.~\eqref{eq:SchwEM1}-\eqref{eq:SchwEMi3} that
satisfy boundary conditions at $r=0$ and $r=\infty$.  At $r=0$ one
imposes the regularity condition that the $q_{(i)}(0)$, $i=0,1,2,3$,
are finite. At $r=\infty$ one imposes
a reflective boundary condition, i.e.,  the $q_{(i)}$
are zero, $q_{(i)}(\infty) \rightarrow 0$.

The solution for $q_{(0)}$ follows either from Eq.~\eqref{eq:SchwEM1}
or from the Bianchi identity Eq.~\eqref{eq:LorenzConditionSchwq} once
one knows the solutions for $q_{(1)}$ and $q_{(2)}$. Thus, in this
sense, one needs only to find the solutions for $q_{(1)}$, $q_{(2)}$,
and $q_{(3)}$, and then $q_{(0)}$ is directly found from $q_{(1)}$ and
$q_{(2)}$. For $i=1,2,3$ we can write concisely
Eqs.~\eqref{eq:SchwEMi}-\eqref{eq:SchwEMi3} as
$\hat{\mathcal{D}}_{j_i} q_{(i)} = 0$, where again $j_1 = \ell -1$,
$j_2 = \ell + 1$, $j_3 = \ell$, and ${\mathcal{D}}_{j_i}$ is defined
in Eq.~\eqref{D}. The solutions of
Eqs.~\eqref{eq:SchwEMi}-\eqref{eq:SchwEMi3}, i.e., of
$\hat{\mathcal{D}}_{j_i} q_{(i)} = 0$, $i=1,2,3$, can be expressed
analytically, as we shall see.  For that we assume that the
time dependence of the
$q_{(i)}$ is $e^{-i\omega t}$, where $\omega$ is the frequency.  Now,
making the change of variables $r_* = l
\arctan\left(\frac{r}{l}\right)$, one finds that
$\hat{\mathcal{D}}_{j_i} q_{(i)} = 0$, $i=1,2,3$, becomes
\begin{align}
   q_{(i)}''
    + \left[\omega^2
    - \frac{j_i(j_i + 1)}{l^2\sin^2\left(\frac{r_*}{l}\right)}
    - \frac{\mu^2}{\cos^2\left(\frac{r_*}{l}\right)}\right]
   q_{(i)} 
  = 0\,.\label{eq:adsnormal1}
\end{align}
The solution of Eq.~\eqref{eq:adsnormal1}
can be worked out, see Appendix \ref{derivation},
to give
that the  
normal mode eigenfunctions $q_{(i)}$, $i=1,2,3,$ have the form
\begin{align}
q_{(i)} =& A_{(i)}\left(\frac{r}{l}\right)^{j_i + 1}
\left(1 + \left(\frac{r}{l}\right)^{2}\right)^{n-
\frac{\omega_i l}{2}} \nonumber\\
&\times \prescript{}{2}{F_1}\left[\omega_i l - n,-n,j_i
+ \frac{3}{2};\frac{\left(\frac{r}{l}\right)^2}{1
+\left(\frac{r}{l}\right)^2}\right]
\label{functionsfinalqi}
\,\,,
\end{align}
where $\prescript{}{2}{F_1}$ is the Gaussian hypergeometric function,
and the frequencies $\omega_i$ corresponding to each $q_{(i)}$ must
obey
\begin{align}
  \omega_i l
  = 2 n + j_i + \frac{3}{2}
  + \frac{1}{2}
  \sqrt{1 + 4 \mu^2 l^2}
  \,.
\label{eq:normalmodesAds}
\end{align}
These are the frequencies for the normal modes of the Proca field in
AdS. Note the lowest frequency corresponds to $j_1 = 0$, i.e., $\ell =
1$ for $q_{(1)}$. The frequencies corresponding to the monopole case
$\ell = 0$ are $\omega_2 l$ with $j_2 = 1$.

The normal modes in~\cite{Konoplya:2006}, Eq.~(44), for the monopole
case $\ell = 0$ do not agree with our expression.  We think this has
to do with a discrepancy between our expression of $\beta$ and the
corresponding expression given in~\cite{Konoplya:2006}. Additional
evidence for the correctness of the present form of the expression is
given by the numerical calculations we have performed to verify the
analytical expression.

The expression of the normal mode frequencies of the Proca field in
AdS given in Eq.~\eqref{eq:normalmodesAds} differs from the expression
given in \cite{ovsiyuk2011spherical} in the case of $j_1 = 0$, where
their Eq.~(24) has an additional term $+1$ compared with our
Eq.~\eqref{eq:normalmodesAds}, which is perhaps simply a typo. Note
that the method used in \cite{ovsiyuk2011spherical} is completely
different from the VSH method we use. Additional evidence for the
correctness of our result is given by the numerical calculations we
have performed to verify the analytical expression. These numerical
calculations were done by using {\it Mathematica}. In particular, 
we used the function {\it NDEigenvalues} to obtain the eigenfrequencies
of Eq.~\eqref{eq:adsnormal1}.

\subsection{Normal modes of the Maxwell field
as the zero mass limit of the Proca field}

The functions and the frequencies of the normal modes for the Maxwell
field can be taken with care by setting~$\mu = 0$ in
Eqs.~\eqref{functionsfinalqi} and~\eqref{eq:normalmodesAds}.  In fact,
due to the gauge freedom, one has that the physical functions
are~$q_{(12)}$ and~$q_{(3)}$ as stated in Eqs.~\eqref{q12Max}
and~\eqref{q3Max} and both have the same normal mode frequency
$\omega$, i.e., $\omega\equiv\omega_{(12)} =\omega_{(3)}$.  The
function~$q_{(12)}$ has a complicated form and we do not show it.  The
function~$q_{(3)}$, let us denote simply by~$q$ in the Maxwell case,
has the form
\begin{align}
q = &A\left(\frac{r}{l}\right)^{\ell + 1}
\left(1 + \left(\frac{r}{l}\right)^{2}
\right)^{n-\frac{\omega l}{2}} \nonumber\\
&\times \prescript{}{2}{F_1}\left[\omega l -
n,-n,\ell + \frac{3}{2};\frac{
\left(\frac{r}{l}\right)^2}{1+\left(
\frac{r}{l}\right)^2}\right]\,\,,
\end{align}
and the frequencies~$\omega$ corresponding to~$q$ must obey
\begin{align}
\omega l = 2n + \ell + 2
  \,.\label{eq:normalmodesAdsMax}
\end{align}

The expression of the normal mode frequencies of the Maxwell field in
AdS given in Eq.~\eqref{eq:normalmodesAds} is the same as that given
in~\cite{Cardoso:2003cj}. Additional evidence for the correctness of
the result is again given by the numerical calculations we have
performed to verify the analytical expression.

\section{Conclusions \label{sec:Conc}}

In this work, the normal modes of the Proca field were obtained
analytically for the AdS spacetime using a
different and new manner, specifically, implementing a
VSH decomposition in a
basis that the decoupling of the equations is direct.
The normal mode
frequencies in the monopole case calculated by us differ from those
found in~\cite{Konoplya:2006}. The normal mode frequencies we found
coincide with the energy levels calculated
in~\cite{ovsiyuk2011spherical}, except for the single case of~$j_1 =
0$. The normal modes of the Maxwell field, as the zero mass limit of
the Proca field, were obtained analytically.
It was shown
that this limit has to be taken
with care in order that one gets rid of the nonphysical gauge modes.

\acknowledgments

This work was supported through the European Research Council
Consolidator Grant 647839, the Portuguese Science Foundation FCT
Project~IF/00577/2015, the FCT Project~PTDC/MAT-APL/30043/2017, the
FCT Project~No.~UIDB/00099/2020, and the FCT
Project~No.~UIDP/00099/2020.
This work has received funding from the European Union's Horizon 2020
research and innovation programme under the Marie Sklodowska-Curie
grant agreement No.~101007855.
The authors would like to acknowledge networking support by the
GWverse COST Action CA16104, ``Black holes, gravitational waves and
fundamental physics.''


\vskip 4.0cm

\appendix

\section{Notation}
\label{notation}

The ansatz used here to decouple and separate the Proca equations,
see Eqs.~\eqref{eq:SchwParamq} and \eqref{Qs}, can
be turned into the notation of \cite{Rosa:2012}. In~\cite{Rosa:2012}
$A_{a}$ is expanded as
\begin{align}
  A_{a} = \frac{1}{r}\sum_{i=0}^3\sum_{l m} c_i u_{(i)}(t,r)
  Z_{(i)a}^{\ell m}(r,\theta,\phi)\,,
  \label{eq:SchwParamA}
\end{align}
where~$c_0 = $, $c_1 = 1$, $c_2 = (\ell(\ell+1))^{-\frac{1}{2}}$, $c_3
= (\ell(\ell+1))^{-\frac{1}{2}}$, and the~$Z_{a}^{(i)\ell m}$ are
given by
\begin{align}
    Z_{(0)a}^{\ell m} &= (1,0,0,0) Y^{\ell m}\,,\nonumber\\
    Z_{(1)a}^{\ell m} &= (0,\dfrac1f,0,0)Y^{\ell m}\,,\nonumber\\
    Z_{(2)a}^{\ell m} &= \frac{r}{\sqrt{\ell(\ell+1)}}
    (0,0,\p_{\theta},\p_{\phi})Y^{\ell m}\,,\nonumber\\
    Z_{(3)a}^{\ell m} &= \frac{r}{\sqrt{\ell(\ell+1)}}
    \Bigg(0,0,\frac{\p_{\phi}}{\sin\theta},
    -\sin\theta\p_{\theta}\Bigg)Y^{\ell m}\,.
\end{align}
The functions~$u_{(i)}$ should be written as~$u_{(i)}^{\ell m}$ but to
avoid overcrowding the notation we have dropped the indices~${\ell
  m}$. The~$u_{(i)}$ have the following correspondence with
the~$q_{(i)}$
\begin{align}
    &u_{(0)} = q_{(0)}\,\,,\label{eq:Acorr0}\\
    &u_{(1)} = q_{(1)} + q_{(2)}\,\,,\label{eq:Acorr1}\\
    &u_{(2)} = (\ell + 1)q_{(2)} - \ell q_{(1)}\,\,,
    \label{eq:Acorr2}\\
    &u_{(3)} = q_{(3)}\,\,.\label{eq:Acorr3}
\end{align}
The Proca equations with the ansatz Eq.~\eqref{eq:SchwParamA} for the
Proca field~$A_a$ become
\begin{align}
  &\hat{\mathcal{D}}_\ell u_{(0)} +
  (\p_r f)(\dot{u}_{(1)} - u'_{(0)}) = 0 \,,
  \label{eq:SchwEM1a}\\
&  \hat{\mathcal{D}}_\ell  u_{(1)} + \frac{2f}{r^2} (u_{(2)} -
u_{(1)})=0\,,
  \label{eq:SchwEM2bettera}\\
&\hat{\mathcal{D}}_\ell u_{(2)}
  + \Bigg[\frac{2f \ell(\ell+1)}{r^2}u_{(1)}\Bigg] = 0\,,
  \label{eq:SchwEM3a}\\
  &\hat{\mathcal{D}}_\ell u_{(3)} = 0\,,\label{eq:SchwEM4a}
\end{align}
and the Bianchi identity becomes
\begin{equation}
  \frac{1}{r f}\Bigg[u'_{(1)} - \dot{u}_{(0)}
  + \frac{f}{r} (u_{(1)} - u_{(2)}) \Bigg] = 0\,,
  \label{eq:LorenzConditionSchwa}
\end{equation}
which was used to find Eq.~(\ref{eq:SchwEM2bettera}). Indeed, by
putting the ansatz Eq.~(\ref{eq:SchwParamA}) onto
Eq.~\eqref{eq:Procaeqvacuum}, one obtains directly
\begin{align}
  \hat{\mathcal{D}}_\ell u_{(1)} + (\p_r f) (\dot{u}_{(0)} -
  u'_{(1)}) + \frac{2 f^2}{r^2}(u_{(2)} - u_{(1)}) = 0
\end{align}
which upon using Eq.~(\ref{eq:LorenzConditionSchwa}) gives
Eq.~(\ref{eq:SchwEM2bettera}). Note the Eqs.~\eqref{eq:SchwEM2bettera}
and~\eqref{eq:SchwEM3a} can be decoupled with the correspondence in
Eqs.~\eqref{eq:Acorr1} and~\eqref{eq:Acorr2}, obtaining
Eqs.~\eqref{eq:SchwEMi}-\eqref{eq:SchwEMi3}.

\section{Derivation of the
normal mode eigenfunctions and the normal mode
frequencies of the Proca field in AdS spacetime
of Sec.~\ref{sec:normalmodes}}
\label{derivation}

For completeness we repeat here
 the normal mode
equation for the Proca field
eigenfunctions $q_{(i)}$
in AdS spacetime,
i.e., 
Eq.~\eqref{eq:adsnormal1}
of the main text, 
\begin{align}
   q_{(i)}''
    + \left[\omega^2
    - \frac{j_i(j_i + 1)}{l^2\sin^2\left(\frac{r_*}{l}\right)}
    - \frac{\mu^2}{\cos^2\left(\frac{r_*}{l}\right)}\right]
   q_{(i)} 
  = 0\,,\label{eq:adsnormalapp}
\end{align}
where all quantities have been previously defined.
We perform the transformation $z =
\sin^2\frac{r_*}{l}$,
recalling that 
$r_*$ is defined by 
$\dv{r_*}{r} =
\frac1f$ with 
$f(r) = 1 +\frac{r^2}{l^2}$. 
We also 
make an ansatz in the variable $z$ of the type
\begin{align}
q_{(i)} = z^\alpha (1-z)^\beta
\psi_{(i)}
\,,\label{eq:app1}
\end{align}
for some new $\psi_{(i)}$, with $\alpha = \frac{1}{2}(j_i
+ 1)$, and $\beta = \frac{1}{4}\Big[ 1 + \sqrt{1 + 4 \mu^2 l^2}\Big]$.
Then, from Eqs.~\eqref{eq:adsnormalapp} and \eqref{eq:app1}
we find that $\psi_{(i)}$ satisfies a hypergeometric differential
equation, namely,
\begin{align}
z(1-z)\p_z^2 \psi_{(i)} + \Big[c - (a + b + 1)z
  \Big]\p_z \psi_{(i)} - ab \psi_{(i)} = 0
\,,\label{eq:hyper1app}
\end{align}
with $a = \alpha + \beta + \frac{\omega l}{2}$, $b = \alpha + \beta -
\frac{\omega l}{2}$, and $c = \frac{1}{2} + 2\alpha$.  The solutions
of Eq.~\eqref{eq:hyper1app} are described by the hypergeometric
function~$\prescript{}{2}{F_1}$, specifically,
\begin{align}
\psi_{(i)} = A_{(i)}\,& \prescript{}{2}{F_1}[a,b,c;z] +
B_{(i)} \,z^{\frac{1}{2} - 2\alpha}\times
\nonumber\\
&\prescript{}{2}{F_1} [d-c, e - c, 2 - c; z]
\,,\label{eq:hyper2app}
\end{align}
where~$A_{(i)}$ and $B_{(i)}$ are constants of integration, and $d =
1+a$ and~$e = 1+b$. 
Thus, from 
Eqs.~\eqref{eq:app1} and \eqref{eq:hyper2app}
we obtain that each $q_{(i)}$, $i=1,2,3$, is given by
\begin{align}
q_{(i)} = &A_{(i)} z^\alpha
(1-z)^\beta \prescript{}{2}{F_1}[a,b,c;z] +
\nonumber\\
&
B_{(i)} z^{\frac{1}{2} -
  \alpha} (1-z)^\beta \prescript{}{2}{F_1} [d-c, e - c, 2 - c; z]
\,,\label{eq:hyper3app}
\end{align}
The first boundary condition states that the
$q_{(i)}$ must be regular at~$r=0$, i.e., at $z\rightarrow0$.  The
hypergeometric function assumes the value of unity
when~$z=0$. Since~$\alpha \geq \frac{1}{2}$ and $\beta \geq
\frac{1}{2}$, the second part of the solution explodes at $z=0$,
thus~$B_{(i)}$
in Eq.~\eqref{eq:hyper3app} must be set to~$0$.
In brief,
the first boundary condition for the Proca field
eigenfunctions $q_{(i)}$, which is imposed
at the center of AdS spacetime, $r=0$ or $z=0$, yields
\begin{align}
B_{(i)}=0
\,.\label{eq:bc1}
\end{align}
The second boundary
condition
is a reflective boundary
condition, it
states that at infinity, $r=\infty$, the $q_{(i)}$ must be
zero.  Following~\cite{Abramowitz:1964}, at $r=\infty$, i.e., $z=1$,
one has that $\prescript{}{2}{F_1}[a,b,c;1] =
\frac{\Gamma(c)\Gamma(w)}{\Gamma(c - a)\Gamma(c-b)}$, where $\Gamma$
is the gamma function, $w = c - a - b$, and $v = a + b -c$, if $a$,
$b$, and $c$ are positive. The solution explodes at~$z=1$ since~$w =
\frac{1}{2} - 2\beta \leq 0$.  The only way that $q_{(i)}$ vanishes at
infinity is if either~$a$ or~$b$ is a negative integer, say $-n$. Now,
$a$ cannot be negative because we require that the frequency must be
positive, by definition.  Hence, the boundary condition at infinity,
$r=\infty$, imposes that $b = - n$.
In brief,
the second boundary condition for the Proca field
eigenfunctions $q_{(i)}$, which is imposed
at the infinity of AdS spacetime, $r=\infty$ or $z=\infty$, yields
with the help
of the previous definitions the result
\begin{align}
&\alpha=\frac{1}{2}(j_i +1) \,,\quad\quad 
\beta= \frac{\omega_i l}{2} - n -\frac{1}{2}(j_i +1)\,, \quad\quad
\nonumber \\
&a=\omega_i l - n\,,\quad\quad 
b=-n\,,\quad\quad c=j_i
+ \frac{3}{2}\,.
\label{eq:bc2}
\end{align}
Collecting all the results,
namely, using Eqs.~\eqref{eq:hyper3app}-\eqref{eq:bc2},
and recalling 
that $z =
\sin^2\frac{r_*}{l}$,
and $r_*$ is defined by 
$\dv{r_*}{r} =
\frac1f$ with 
$f(r) = 1 +\frac{r^2}{l^2}$, 
we
have that the
normal mode eigenfunctions 
$q_{(i)}$, $i=1,2,3,$ have the form
\begin{align}
q_{(i)} =& A_{(i)}\left(\frac{r}{l}\right)^{j_i + 1}
\left(1 + \left(\frac{r}{l}\right)^{2}\right)^{n-
\frac{\omega_i l}{2}} \times\nonumber\\
& \prescript{}{2}{F_1}\left[\omega_i l - n,-n,j_i
+ \frac{3}{2};\frac{\left(\frac{r}{l}\right)^2}{1
+\left(\frac{r}{l}\right)^2}\right]
\label{functionsfinalqiapp}
\,.
\end{align}
and the frequencies $\omega_i$ corresponding to each $q_{(i)}$ must
obey
\begin{align}
  \omega_i l
  = 2 n + j_i + \frac{3}{2}
  + \frac{1}{2}
  \sqrt{1 + 4 \mu^2 l^2}
  \,.
\label{eq:normalmodesAdsapp}
\end{align}
These are the frequencies for the normal modes of the Proca field in
AdS. Note the lowest frequency corresponds to $j_1 = 0$, i.e., $\ell =
1$ for $q_{(1)}$.
Note that Eqs.~\eqref{functionsfinalqiapp} and
\eqref{eq:normalmodesAdsapp} are Eqs.~\eqref{functionsfinalqi} and
\eqref{eq:normalmodesAds} of the main text.



\end{document}